# Enhancement of optical absorption in multiferroic (1-x)PZT-xPFN thin films: Experiments and first-principles analysis

L. Imhoff, M. B. Di Marco, C. Lavado, S. Barolin and M. G. Stachiotti

*Instituto de Física de Rosario (IFIR), CONICET – UNR,
Bv. 27 de Febrero 210 Bis, (2000) Rosario, Argentina.*

Multiferroic compounds have gained research attention in the field of ferroelectric photovoltaics due to the presence of transition-metal *d* states from magnetic ions, which tend to reduce the bandgap value. In this work, $0.5Pb(Zr_{0.52}Ti_{0.48})O_3$–$0.5Pb(Fe_{0.5}Nb_{0.5})O_3$ [$PZTFN_{0.5}$] thin films were synthesized using a sol-gel route to investigate the effect of iron doping on optical and multiferroic properties. For comparative analysis, the end-member compositions, $Pb(Zr_{0.52}Ti_{0.48})O_3$ [PZT] and $Pb(Fe_{0.5}Nb_{0.5})O_3$ [PFN], were also synthesized under identical conditions. Our results revealed that the presence of $Fe^{3+}$ ions, besides inducing multiferroic behavior, effectively enhances the optical absorption of the material in the visible light region. Optical transitions at ~3.0 eV (~2.4 eV) and ~2.7 eV (~2.2 eV) for the direct (indirect) bandgap were determined for $PZTFN_{0.5}$ and PFN, respectively, indicating that the absorption edges of the iron-containing films result more promising than PZT ($E_g$~3.6eV) for photovoltaic applications. Both $PZTFN_{0.5}$ and PFN thin films exhibit multiferroic behavior at room temperature, with different electric and magnetic properties. While $PZTFN_{0.5}$ presents saturated hysteresis loops with remanent polarization values around 10 $\mu C/cm^2$ and magnetization of 1.6 $emu/cm^2$, PFN displays significantly larger remanence (31 $emu/cm^2$) but poorer ferroelectric properties due to the presence of leakage. Microscopic insights into the structural and electronic properties of the $PZTFN_{0.5}$ solid solution were provided from first-principles calculations.

Keywords: PZT-PFN. Thin films, Multiferroics, Photovoltaics

## I. Introduction

Multiferroic materials have attracted significant attention for both fundamental understanding and potential applications in a wide variety of multifunctional devices. Among them, BiFeO$_3$ (BFO) is the most extensively studied single-phase multiferroic material due to the coexistence of magnetic and ferroelectric orders under ambient conditions. Interestingly, BFO has also been the subject of numerous studies for photovoltaic applications [1–3], due to its high remanent polarization and narrow bandgap in comparison to other ferroelectric materials. In fact, the use of ferroelectric oxides for photovoltaic applications presents a rising interest in the research area [4–8], since these materials can exhibit a very high photovoltage, up to orders of magnitude larger than their bandgap. Many studies were conducted on archetypical ABO$_3$ perovskites such as, e.g., Pb(Zr,Ti)O$_3$ (PZT) [9–13], PbTiO$_3$ (PTO) [14,15], and BaTiO$_3$ (BTO) [16,17]. However, these perovskites present large bandgaps (> 3eV), that limit the absorption of those materials to ultraviolet light, which makes up only around 8% of the solar spectrum. The smaller bandgap of ~2.67 eV that is observed in BFO films [18] enables BFO to absorb a greater number of photons for energy conversion [19,20].

The photocurrent generation in multiferroics is fundamentally similar to the photovoltaic effect in ferroelectrics, that is, the spontaneous polarization is the compelling force for separating photoinduced electrons and holes. However, in multiferroic materials, there would be possibilities of both electric-field and magnetic-field control of properties, such as the photocurrent direction [21]. Nowadays, extensive research is being carried out to find and design novel materials that exhibit both ferromagnetic and ferroelectric order at room temperature [22,23]. Among these, the incorporation of magnetic ($d^5$) Fe$^{3+}$ ions and nonmagnetic ($d^0$) Nb$^{5+}$ (or Ta$^{5+}$) ions sharing the B site of the PZT perovskite structure has proven to be a successful strategy for the development of single-phase multiferroic compounds, both in bulk [24–31] and thin film form [32,33]. Note that

$Fe^{3+}/Nb^{5+}$ co-substitution into the B-site of the perovskite PZT structure avoids conductivity issues, since the generation of oxygen vacancies by the addition of $Fe^{3+}$ would be suppressed for equimolar compositions of +3 and +5 cations. This system, which can be understood as a solid solution between PZT and $PbFe_{0.5}Nb_{0.5}O_3$ (PFN), displays multiferroic behavior at room temperature. In $(1-x)Pb(Zr_{0.52}Ti_{0.48})O_3-xPb(Fe_{0.5}Nb_{0.5})O_3$ bulk ceramics, for instance, magnetization loops were observed for compositions between x=0.1 and 0.4, with an improvement in the magnetization values for x=0.3 [26,31]. Ferroelectric hysteresis loops with remanent polarization values of ~20–30 $\mu C/cm^2$ have been observed [24–26,31]. In thin films, room-temperature ferromagnetic behavior was observed for x>0.3, with saturated ferroelectric hysteresis loops and remanent polarization values of ~12$\mu C/cm^2$ [33]. While the above-mentioned works were all conducted to characterize multiferroic properties, little is known about the effect of the $Fe^{3+}/Nb^{5+}$ co-doping on the electronic properties of the material. In particular, it would be interesting to determine to what extent Fe/Nb co-doping modifies the bandgap of the PZT material to enhance the optical absorption ability. In this work, $0.5Pb(Zr_{0.52}Ti_{0.48})O_3-0.5Pb(Fe_{0.5}Nb_{0.5})O_3$ [PZTFN$_{0.5}$] thin films were synthesized to quantitatively investigate the optical bandgap. PZT and PFN thin films were also fabricated for comparative studies. We show that the addition of $Fe^{3+}$ and $Nb^{5+}$ ions, besides inducing multiferroic behavior, effectively enhances the optical absorption of the material in the visible light region. The Tauc plot analysis reveal optical transitions that result more promising for photovoltaic applications compared to PZT, shifting the absorption onset from the UV to the blue range of the visible spectrum. First-principles electronic structure calculations were performed to provide microscopic insights into the structural and electronic properties of the multiferroic solid solution.

**II. Methods**

The films were synthesized by modified sol-gel routes developed at our laboratory. Two compatible routes, based on the use of acetoin (3-hydroxy-2butanone) as a chelating agent, were developed for the synthesis of PZT and PFN precursors, with concentration values of 0.2 M. The 0.5PZT-0.5PFN precursor was then prepared by mixing equal volumes of PZT and PFN solutions. Details about the preparation process are presented in reference [33]. The films were deposited by spin-coating (4000 rpm, 15 seconds) onto Pt/Ti/SiO$_2$/Si wafers and fluorine-doped tin oxide (FTO) coated aluminum-borosilicate glass substrates. The samples were thermally treated at 650 °C by rapid thermal annealing (RTA) with a heating rate of 50 °C/s, which resulted the best thermal treatment for the fabrication of good-quality PZTFNx thin films. The whole process was repeated six times to obtain a 6-layered film, and the final thicknesses of the sintered films were determined to be around 240 nm.

The films were characterized by X-ray diffraction using a Phillips X'pert Pro X-Ray diffractometer with Cu Kα radiation of wavelength 1.5406 Å. The measurements were taken in the grazing incident configuration with an incident beam angle of 5º, 2θ varying between 20º and 60º and a scanning time of 0.02 º/s. Raman spectra were acquired with a Renishaw inVia Raman spectrometer by means of the 514 nm Ar-ion laser line (10mW nominal power). The surface morphology of the samples was observed by atomic force microscopy (Nanotec ELECTRONIC AFM equipment) in tapping mode using a silicon probe with 150 kHz resonance frequency and a constant force of 7.4 N/m. For electrical measurements, 0.25 mm diameter platinum top electrodes were deposited by DC-sputtering on the surface of the films, and then thermally treated at 350 °C for 10 minutes. Dielectric properties were measured using an impedance meter (LCR meter) QuadTech 7600 plus. The ferroelectric cycles were obtained at room temperature from a Sawyer-Tower circuit applying alternate signals at frequencies of 50 Hz and 1 kHz. The magnetic measurements were carried out in a commercial superconducting quantum interference device magnetometer (SQUID). UV-visible light transmission and absorption characteristics were measured by a JASCO UV/VIS 550.

For the first-principles electronic structure calculations we used the plane augmented wave method (PAW) as implemented in the Vienna Ab-Initio Package (VASP) [34,35], selecting the Perdew–Burke–Ernzerhof form of the exchange-correlation functional for solids (PBEsol). For structural relaxations we considered a cutoff of 600 eV, ensuring a convergence of 0.005 eV/Å for the residual ionic forces. The U parameter was employed on Fe sites only, and was initially set at 4 eV (later it was varied to see the difference). The k-mesh and supercell size were $4 \times 4 \times 4$ and $2 \times 2 \times 2$, respectively, and included two Fe and two Nb ions.

## III. Experiments

The films were deposited onto Pt/Ti/SiO$_2$/Si wafers for electric and magnetic characterizations, and onto FTO-coated aluminum-borosilicate glass (FTO-AG) substrates for optical measurements. The latter substrate is adequate for the deposition of PZT films due to its high thermal resistance and low cost compared to quartz [36].

The X-ray diffraction patterns of the synthesized films, shown in Fig. 1, indicate that polycrystalline thin films with perovskite structure were successfully grown on both substrates. Pyrochlore-free films are obtained on platinized silicon substrates (left panel), whereas on FTO-AG substrates, the deposition of iron-containing films reveals some pyrochlore phase formation (right panel). The broad shape of the (002) and (112) peaks suggest the presence of tetragonal symmetry in PZT and PZTFN$_{0.5}$ films deposited on both kinds of substrates. On the other hand, these peaks become sharper for PFN films, indicating a pseudocubic structure [37,38].

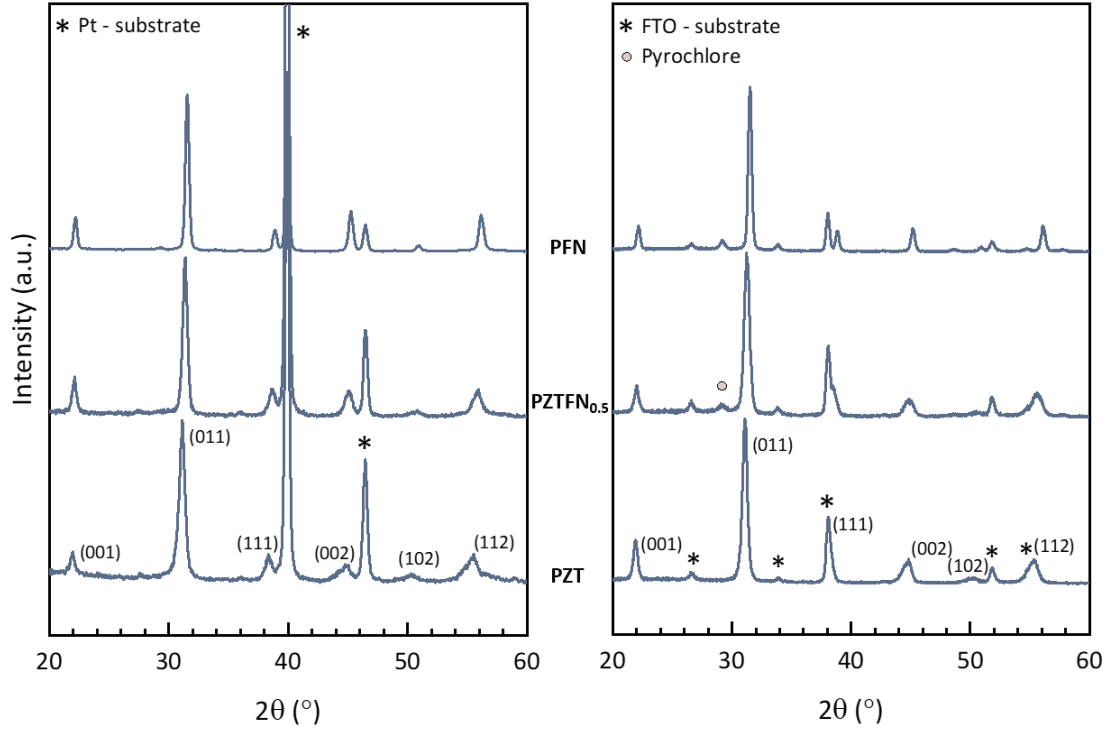

**Figure 1:** XRD patterns of PZT, PZTFN$_{0.5}$ and PFN films deposited onto **(a)** Pt/Ti/SiO$_2$/Si buffers and onto **(b)** FTO-coated aluminum-borosilicate glass substrates. The films were annealed at 650 °C by RTA. The corresponding perovskite peaks are indexed.

| Sample | a (Å) | c (Å) | c/a | V (Å$^3$) | Sigma | R$_{rw}$ (%) |
|---|---|---|---|---|---|---|
| **PZT** | 4.034 | 4.094 | 1.015 | 66.62 | 1.61 | 21 |
| **PZTFN$_{0.5}$** | 4.020 | 4.053 | 1.008 | 65.50 | 1.55 | 16 |
| **PFN** | 3.997 | 4.012 | 1.004 | 64.10 | 1.21 | 13 |

**Table 1:** Cell parameters calculated for the three compositions deposited on Pt/Ti/SiO2/Si substrates.

A Rietveld refinement was performed for the films deposited on platinized substrates assuming a tetragonal P4mm lattice geometry, using the software MAUD [39]. The lattice parameters obtained, along with their corresponding quality factors, are presented in Table 1. The analysis confirmed a reduction in tetragonality as the iron content in the samples increases. Additionally, cell volumes were calculated, showing a decrease in their values with the increase in iron content.

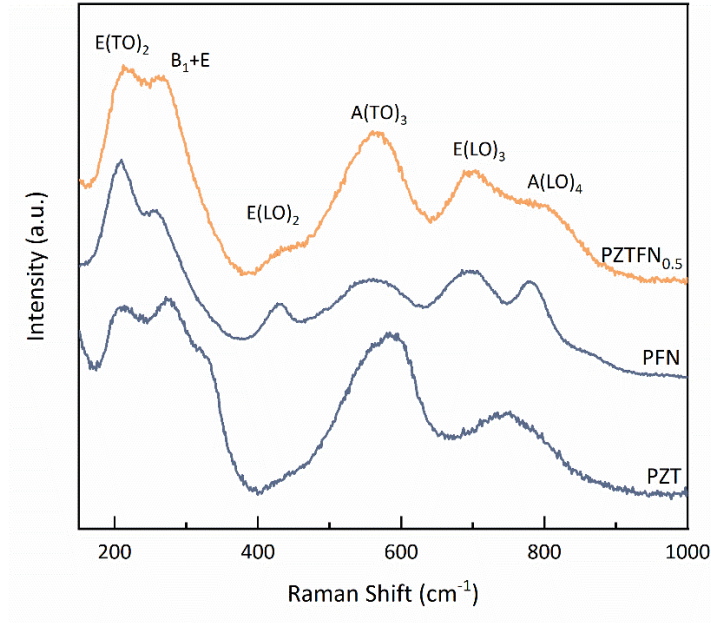

**Figure 2:** Room temperature Raman spectra for PZT, PZTFN$_{0.5}$ and PFN thin films.

Raman spectroscopy was employed to further study the structural properties of the three compositions. Figure 2 shows the Raman spectra measured at room temperature for PZT, PZTFN$_{0.5}$ and PFN thin films. The spectra for PZT and PFN contain broad overlapping bands, which agree with those found in the literature [40,41]. According to the factor group analysis, tetragonal PZT has 12 optical normal modes of symmetry that are Raman-active modes: $4E+3A_1+B_1$. When describing the Raman spectra of PFN and other pseudocubic relaxor ferroelectrics, the Fm-3m symmetry is considered a good approximation [41,42]. For this symmetry, group theory predicts four Raman active modes $A_{1g}+E_g+2F_{2g}$, and the $4F_{1u}$ infrared (IR)-active modes. Note, however, that ionic disorder and off-center displacements lead to a local symmetry breaking, and the spectra contain more lines than expected for the ideal symmetry of the crystal structure. Figure 2 shows that the spectrum for PZTFN$_{0.5}$ can be assumed as a superposition of PZT and PFN spectra, as it contains bands that can be associated to both of them. For the description of this spectrum, a tetragonal symmetry was considered. The spectrum showed low-frequency phonon modes E(TO$_2$) and E+B$_1$ at ~210 cm$^{-1}$ and 265 cm$^{-1}$, which are attributed to the BO$_6$ bending and stretching vibration, respectively [26]. Four additional peaks appeared at 429, 565, 700 and 780 cm$^{-1}$ that correspond to E(LO$_2$), A(TO$_3$), E(LO$_3$) and A(LO$_4$) modes, respectively. The modes at 429 cm$^{-1}$, 565 cm$^{-1}$ and 694 cm$^{-1}$ can be ascribed to Fe–O stretching vibration, O–B–O bending and B–O stretching of the oxygen octahedron, respectively [43,44]. The A(LO$_4$) mode is generally assigned as the Nb–O–Fe stretching mode [26].

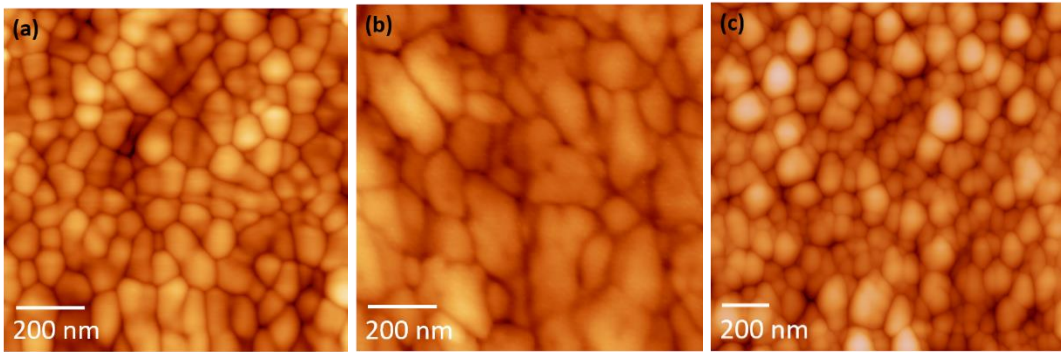

**Figure 3:** Topographic AFM images of **(a)** PZT, **(b)** PZTFN$_{0.5}$ and **(c)** PFN films deposited onto Pt/Ti/SiO$_2$/Si substrates.

AFM images are shown in Fig.3 to visualize the surface morphology of the films deposited on platinized substrates. The observed surfaces are smooth and free of cracks for all three compositions. Both PZT and PFN samples exhibit similar grain shapes and sizes. In the case of the PZTFN$_{0.5}$ sample, grain size values show more dispersion and appear to be larger compared to the other samples. This difference could be attributed to the agglomeration of smaller grains. The RMS roughness was evaluated using equivalently sized images, revealing that PZT and PZTFN$_{0.5}$ have similar values (~15 nm). Interestingly, the RMS roughness for PFN is roughly twice as large (31 nm).

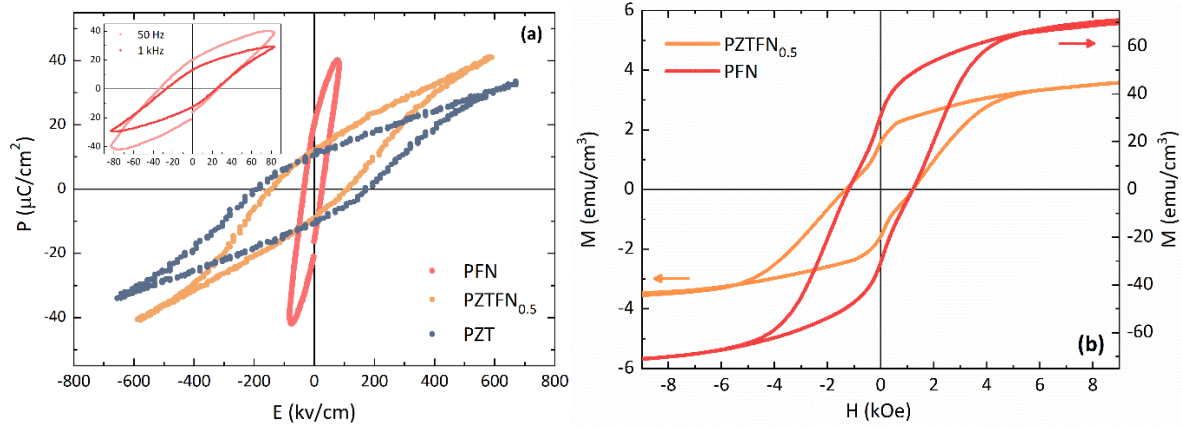

**Figure 4: (a)** Ferroelectric hysteresis loops measured at 50 Hz for PZT, PZTFN$_{0.5}$ and PFN thin films. The loop measured at 50 Hz for the PFN thin film presents signatures of conductivity, therefore a measurement at 1 kHz is also presented in the inset. **(b)** Magnetization loops for PZTFN$_{0.5}$ and PFN films deposited onto Pt/Ti/SiO$_2$/Si buffers.

The ferroelectric and ferromagnetic features of the films are presented in Fig. 4. Figure 4a shows ferroelectric hysteresis loops for the three compositions measured at 50 Hz. PZT and PZTFN$_{0.5}$ films display similar ferroelectric behavior, with well saturated hysteresis loops, remanent polarization (P$_r$) values of ~10 μC/cm$^2$ and coercive fields (E$_c$) of 200 and 120 kV/cm, respectively. The loop measured at 50 Hz for the PFN film is not saturated and displays a small coercive field (~20 kV/cm) with a P$_r$ value of ~20 μC/cm$^2$. However, it's important to note that this value depends on the applied voltage frequency and it becomes lower for higher frequencies. This frequency dependency can be clearly seen in the inset of Figure 4a, where loops measured at 50 Hz and 1 kHz for PFN are compared. We thus conclude that in comparison with the other two samples, the PFN film exhibits poorer ferroelectric properties, mainly due to the presence of leakage, especially at low frequencies. This deduction is based on the observation of low-saturated loops and breakdown electric fields 10 times smaller than the ones obtained for PZTFN$_{0.5}$ and PZT. The presence of unbalanced charges due to nonstoichiometric amounts of B-site cations in PFN, which promotes the presence of oxygen vacancies, can be a contributing factor to the increase in leakage current. A high leakage current is detrimental to the ferroelectric properties [37,45,46].

In Figure 4b, we present the magnetization-field response for the iron-containing films. The M(H) loops clearly indicates the presence of weak ferromagnetism at room temperature. Although the coercive field is nearly identical in both films, the remanence is significantly larger in PFN (~20 times greater). The weak ferromagnetic behavior that is observed in PZTFN$_x$ solid solutions may arise from the clustering of the FeO$_6$ octahedra, forming iron-rich nanoregions [31,47]. Another possible source is the Dzyaloshinskii–Moriya (DM) interaction between Fe$^{3+}$ ions through oxygen ions, which can drive a small canting between the interacting moments. However, the origin of the significantly higher remanence observed in the PFN film remains unclear. It's worth noting that room-temperature weak ferromagnetic behavior has also been observed in PFN bulk ceramics [48–50] and epitaxial films [51,52], and its origin is still subject of discussion.

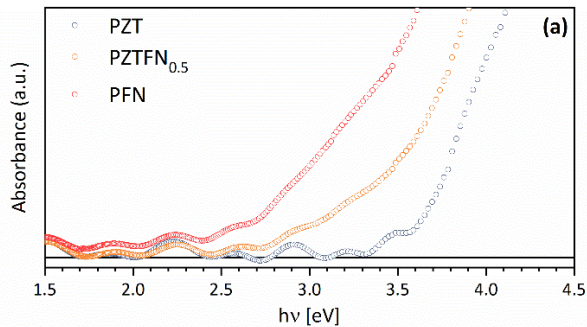
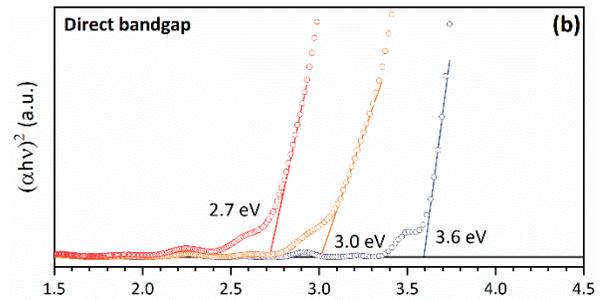
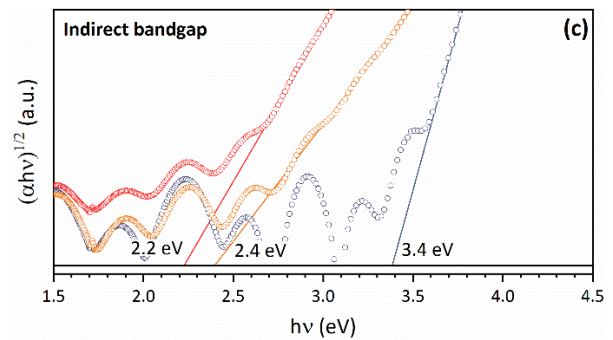

**Figure 5:** **(a)** Absorbance spectra for PZT, PZTFN$_{0.5}$ and PFN films deposited onto FTO-coated aluminoborosilicate glass substrates from UV-Vis absorption measurements in the wavelength range 200–1100 nm. Tauc plots extrapolation for direct **(b)** and indirect **(c)** bandgaps.

We now present the effects of iron content on the optical properties. UV-Vis spectroscopy measurements were carried out on the films deposited on FTO-AG transparent substrates. Figure 5a shows the absorbance spectra obtained at room temperature within the wavelength range of 325 to 1100 nm. The fringes observed in the spectra result from interference between interfaces. The three films display clearly different absorption edges. Increasing the Fe-content shifts the absorption onset towards higher wavelengths, which is an outcome of the increasing ability of the multiferroic films to absorb lower-energy photons. This result suggests a reduction in the bandgap value due to the incorporation of $Fe^{3+}$ and $Nb^{5+}$ ions into the PZT lattice.

For the determination of bandgap energies ($E_g$), the absorption coefficient ($\alpha$) was evaluated from the corresponding absorbance spectra as $\alpha = (2.303\ A) / d$, where *A* is the absorbance and *d* is the film thickness. The gaps were estimated from Tauc plots $(\alpha h\nu)^n$ vs. $h\nu$, where the n factor depends on the nature of the electron transition and is equal to 2 or 1/2 for the direct and indirect transitions, respectively. Here, $h\nu$ results the energy of the incident photon. The $E_g$ values were determined by extending the linear region of the Tauc plots to the horizontal axis. Direct bandgap energies of 3.6, 3.0 and 2.7 eV were obtained for PZT, $PZTFN_{0.5}$ and PFN, respectively (Fig.5b). The bandgap undergoes a redshift with the addition of Fe ions, and we note that the value of 2.7 eV obtained for the PFN film is in good agreement with the one reported for epitaxial thin films synthesized by pulsed laser deposition [53]. This value is also comparable with the data reported for bulk ceramics prepared by mechanochemical activation-assisted synthesis, where a bandgap value of 2.55 eV was measured [54]. The x=0.5 solid solution presents an intermediate bandgap of 3.0 eV. Figure 5c shows indirect transitions at 3.4, 2.4 and 2.2 eV for PZT, $PZTFN_{0.5}$ and PFN, respectively. These transitions can be generated by the presence of indirect gaps or by the formation of defects. An indirect transition at 2.25 eV was observed in epitaxial PFN films [53], and the authors concluded that this absorption is likely to be originated from defect levels in the bandgap. So, bandgaps in PZTFN$_{0.5}$ and PFN are much smaller compared to PZT and the absorption onset in the multiferroic thin films shifts from the UV to the blue range of the visible spectrum, opening up new functionalities for them. PFN absorbs more light in the visible band than PZTFN$_{0.5}$, but its poor ferroelectric properties would be a serious drawback for photovoltaic applications. The multiferroic solid solution, on the other hand, displays good ferroelectric properties with a remanent polarization of ~10 μC/cm$^2$, weak ferromagnetic behavior at room temperature, and a redshift of the optical absorption edge of ~130 nm with respect to PZT.

**IV. Simulations**

Although perovskites in the ternary system PT-PZT-PFN have attracted significant interest due to their ability to exhibit simultaneous ferroelectric and ferromagnetic behavior, there is a lack of first-principles theoretical studies in this system. Here, we used the VASP package for conducting first-principles electronic structure calculations in order to get microscopic insights into the crystal and electronic structure of the PZTFN$_{0.5}$ solid solution. For the sake of simplicity, we first studied the structural properties of the solid solution between PFN and lead titanate (PT): 0.5PbTiO$_3$–0.5Pb(Fe$_{0.5}$Nb$_{0.5}$)O$_3$ [PTFN$_{0.5}$]. A 2 × 2 × 2 supercell with 2Fe, 2Nb and 4Ti was used, and we tried all possible configurations for the transition ions. Full structural relaxation was allowed for all inequivalent configurations. We show in Fig. 6a a bar plot for the calculated energies assuming both ferro (FM) and antiferromagnetic (AFM) coupling between Fe$^{3+}$ ions. The minimum energy corresponds to the arrangement Fe$_{12}$Nb$_{78,}$ with antiparallel alignment of the Fe's magnetic moments. In this configuration, Fe atoms are located at neighboring sites 1 and 2, Nb at sites 7 and 8, and Ti at the rest (see the inset for the reference). The crystal structure of the

Fe$_{12}$Nb$_{78}$ ground state displays tetragonal symmetry, with Fe ions along the c-axis (see Fig.6b). It's worth noting that experiments also suggest a tetragonal ground state for the PT-PFN$_{0.5}$ solid solution [55–57]. The optimized lattice parameters are a=3.926 Å and c=4.107 Å (c/a=1.046). These values compare well with experimental data reported for PTFN$_{0.4}$ solid solutions (a=3.909 Å, c=4.078 Å, c/a=1.042 [57]).

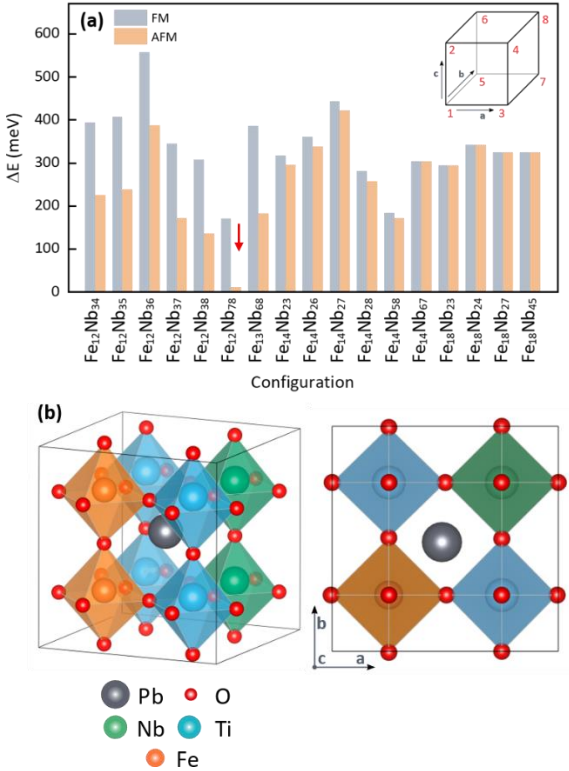
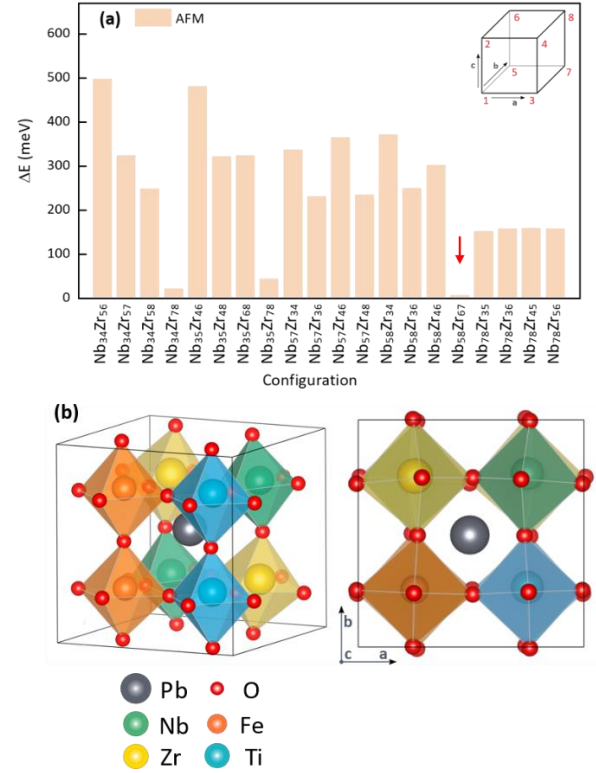

**Figure 6:** **(a)** Relative energies of all chemical configurations for the solid solution 0.5PbTiO$_3$–0.5Pb(Fe$_{0.5}$Nb$_{0.5}$)O$_3$ [PT-PFN$_{0.5}$]. A 2 × 2 × 2 supercell with 2 Fe, 2 Nb and 4 Ti was used for the study, and all configurations were fully relaxed. The inset shows the numbers to identify the corresponding cells. FM (gray bars) and AFM (orange bars) coupling between Fe$^{3+}$ ions are compared. **(b)** Detailed crystal structure of the lowest energy configuration Fe$_{12}$Nb$_{78}$. The structure has tetragonal symmetry with optimized lattice parameters a=3.926 Å and c=4.107 Å.

**Figure 7:** **(a)** Relative energies of the chemical configurations discussed in the main text for the solid solution 0.5PbZr$_{0.5}$Ti$_{0.5}$O$_3$ – 0.5Pb(Fe$_{0.5}$Nb$_{0.5}$)O$_3$ [PZTFN$_{0.5}$]. We used a 2 × 2 × 2 supercell with 2 Fe, 2 Nb, 2 Zr and 2 Ti. We consider all possible configurations with Fe$^{3+}$ ions AFM-coupled occupying sites 1 and 2. The inset shows the numbers to identify the corresponding cells. **(b)** Detailed crystal structure of the lowest energy configuration. The structure has monoclinic symmetry with lattice parameters a=3.985 Å, b=4.040 Å, c=4.055 Å, α=89.99°, β=89.50° and γ=89.98°.

The structural analysis for the $0.5PbZr_{0.5}Ti_{0.5}O_3$–$0.5Pb(Fe_{0.5}Nb_{0.5})O_3$ [$PZTFN_{0.5}$] solid solution was also performed using a 2 × 2 × 2 supercell (2Fe, 2Nb, 2Ti and 2Zr). In this case, we calculated all possible chemical configurations with the constraint of antiferromagnetically coupled Fe ions occupying neighboring sites (sites 1 and 2). Figure 7a shows the energies of the inequivalent configurations. As it can be seen, the energy differences between different cation configurations are similar to those shown in Fig.5a, all below ∼50 meV per unit cell. The lowest energy configuration corresponds to the $Nb_{58}Zr_{67}$ arrangement. There are, however, two other close-energy configurations ($Nb_{34}Zr_{78}$ and $Nb_{35}Zr_{78}$), and these three configurations may compete with each other. Interestingly, the three low-energy structures have monoclinic symmetry (not tetragonal, as in $PTFN_{0.5}$). The crystal structure of the $Nb_{58}Zr_{67}$ arrangement is depicted in Fig.7b (lattice parameters a=3.985 Å, b=4.040 Å, c=4.055 Å, α=89.99°, β=89.50° and γ=89.98°, unit-cell volume 65.28 Å$^3$). Note that the monoclinic symmetry of the ground-state structure predicted by the DFT calculations agrees with experiments. In fact, Schiemer et.al. [28] constructed a phase diagram for the $PZTFN_x$ solid solution from different experimental data, showing that the x=0.5 composition displays cubic (Pm3m) - tetragonal (P4mm) - monoclinic (Cm) transitions with decreasing temperature [28]. We also calculated the structure of the tetragonal phase. For that purpose, we optimized the three competing chemical configurations ($Nb_{58}Zr_{67}$, $Nb_{34}Zr_{78}$ and $Nb_{35}Zr_{78}$) by imposing tetragonal symmetry. We found that the lowest energy configuration corresponds again to the $Nb_{58}Zr_{67}$ arrangement (a=3.988 Å, c=4.136 Å, c/a=1.037), with a configurational energy 8.2 meV/f.u. above the monoclinic ground state. We note that this energy difference compares well with the one observed between the tetragonal and rhombohedral phases in $BaTiO_3$ (~7 meV/f.u using the PBEsol functional [58]), a material that displays a rhombohedral ground-state and a room-temperature tetragonal phase.

We note that the DFT results support the idea of Fe clustering as there is a preference for two $Fe^{3+}$ ions to occupy adjacent sites. This feature, also observed in DFT calculations for PFN [59], indicates that the observed room-temperature FM behavior could be related to the existence of Fe-rich nanoregions, as was recently proposed for $PZTFN_x$ solid solutions with x>0.3 [31]. Within the cluster configuration, and despite the fact that Fe-Fe magnetic interactions are AFM, uncompensated Fe magnetic moments may lead to a net magnetization superimposed on the AFM order [60]. The calculations also reveal the presence of oxygen-octahedron rotations in $PZTFN_{0.5}$ (compare, for instance, the top views shown in Figs. 6b and 7b), which can drive a small canting between interacting moments inducing weak ferromagnetic behavior via the Dzyaloshinsky-Moriya interaction.

In Figure 8a, electronic densities of states (DOS), both total and projected on selected atomic and orbital characters, are shown for the lowest energy chemical configuration $Nb_{58}Zr_{67}$ in the monoclinic phase. A band gap of 1.9 eV is obtained for $U_{Fe} = 4$ eV, the value used throughout this paper. The valence band is formed by states with predominant O-p character while the conduction band is formed by states with predominant transition metal-d character (and some Pb-p character at high energies). The broad peak centered at -7.5 eV corresponds to the Pb-6s band, which is positioned below the O-2p valence band in lead-based $ABO_3$ perovskites (this band is usually interpreted as the $6s^2$ lone pair electrons of Pb). The double-peak structure between the Pb-6s and O-2p bands corresponds to occupied Fe-3d states, while unoccupied Fe-3d states form a shallow mid-gap state at the bottom of the conduction band. The strong splitting of ~8 eV between occupied and unoccupied Fe-d states can be better visualized in Fig.8b, where the spin channels of the projected DOS for the two inequivalent Fe ions are presented. These results confirm the antiferromagnetic coupling between $Fe^{3+}$ ions. The calculated effective magnetic moment of each $Fe^{3+}$ ion is 4.06 $\mu_B$.

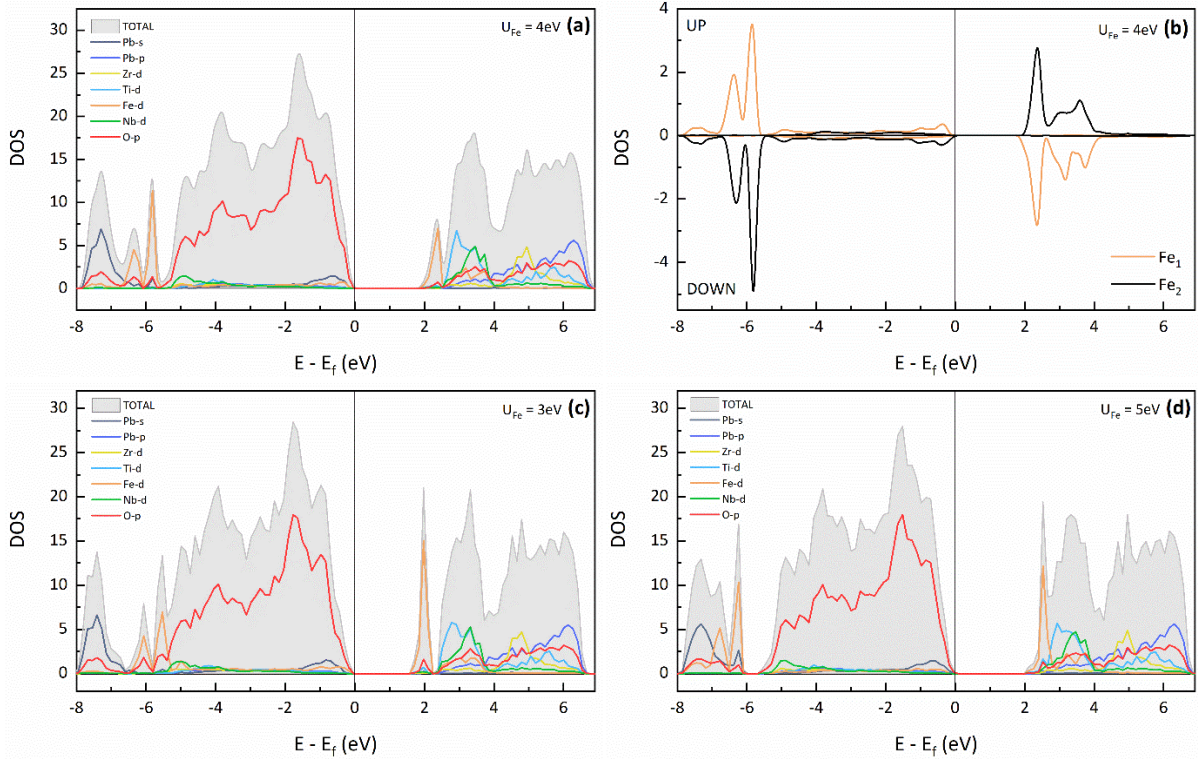

**Figure 8:** **(a)** Total (gray shaded) and projected (lines) densities of states (DOS) for the lowest energy chemical configuration $Nb_{58}Zr_{67}$ in the monoclinic phase using the Hubbard U parameter $U_{Fe}$ = 4.0 eV. **(b)** Spin-resolved DOS for the Fe 3d states of the two inequivalent magnetic ions in the 2x2x2 supercell. The electronic configurations are compatible with $Fe^{3+}$ valence states. **(c)** DOS for the lowest energy chemical configuration $Nb_{58}Zr_{67}$ in the monoclinic phase using $U_{Fe}$ = 3.0 eV and **(d)** $U_{Fe}$ = 5.0 eV.

To test the impact of the $U_{Fe}$ parameter on the electronic structure, we compare the DOS obtained for lower (3 eV) and higher (5 eV) values of $U_{Fe}$. For $U_{Fe}$ = 3 eV (Fig. 8c), the spin splitting of the Fe-d states decreases, the occupied Fe-d states display more hybridization with O-2p states at the bottom of the valence band and the unoccupied states are pushed to lower energies. As a result, the bandgap decreases to 1.6 eV. On the other hand, the larger $U_{Fe}$ value increases the spin splitting of the Fe-d states (Fig. 8d). The occupied Fe-d states split from the O-2p valence band and hybridize with Pb-s band. The unoccupied states are pushed to higher energies, increasing the band gap to 2.1 eV. So, different $U_{Fe}$ parameters slightly modify the bandgap of the system, but in all cases the lower conduction band edge is dominated by Fe-d states. Unfortunately, there is no experimental data available for PZTFN$_x$ solid solutions to identify the orbital character of the conduction band edge. Such data would be useful to determine the specific value for $U_{Fe}$ that must be used in the calculations.

Finally, it is interesting to note that Fe/Nb co-doping induce a reduction of the unit-cell volume, which increases the band widths and reduces bandgap. We performed DFT calculations for PbZr$_{0.5}$Ti$_{0.5}$O$_3$ using a 2 × 2 × 2 supercell with 4 Zr and 4 Ti, and all possible configurations were tested. We found that the lowest energy configuration corresponds to Zr ions located at sites 1, 2, 7, and 8 (according to the numbering shown in the inset of Fig. 7), which is in agreement with another DFT study [61]). The unit-cell volume of this configuration is 67.60 Å$^3$. If Zr and Ti atoms were set in alternating positions, energy minimization leads to an equilibrium volume of 66.43 A$^3$. In both cases, the unit-cell volume is larger than the one obtained for the PZTFN$_{0.5}$ solid solution (65.28 A$^3$). The experimental results shown in Table I supports the DFT results. Bandgaps of ~2.3 eV were obtained for the two chemical configurations of PZT, a larger value than the ones obtained for PZTFN$_{0.5}$ using different values of $U_{Fe}$.

**Conclusions**

0.5PZT-0.5PFN thin films were synthesized by a sol-gel route to investigate their optical and multiferroic properties. The films display single-phase perovskite structure with multiferroic behavior at room temperature. The incorporation of Fe$^{3+}$ and Nb$^{5+}$ ions into PZT induces ferromagnetic behavior, still maintaining a spontaneous ferroelectric polarization similar to the PZT host lattice. UV-Vis spectroscopy measurements revealed a redshift for the optical absorption edge, ~130 nm above the edge of the PZT film. Tauc plot analysis indicate optical transitions at ~3.0 eV and ~2.4 eV for the direct and indirect bandgap, respectively. Structural and electronic features were provided by first-principles calculations. The simulations support the idea of Fe clustering with spin-canting, indicating that the weak ferromagnetic behavior observed at room temperature could be related to the existence of Fe-rich nanoregions. The calculations also revealed a reduction in the unit-cell volume upon Fe doping and the presence of Fe-3d mid-gap states in the bandgap, which are in accordance with the observed redshift. We have also showed that the visible range optical absorption in PFN thin films is larger than in $PZTFN_{0.5}$, with optical transitions at ~2.7 eV and ~2.2 eV for the direct and indirect bandgap, respectively. However, the poor ferroelectric properties of PFN films prevent its use as a photovoltaic material. The reduction of the optical bandgap without compromising the good ferroelectric properties makes the $PZTFN_x$ solid solution a promising candidate for ferro-photovoltaic applications.

**Acknowledgements**

The authors thank Myriam H. Aguirre and Miguel A. Rengifo for the access to equipment at INMA- University of Zaragoza to perform the electric and magnetic measurements.

**Funding**

This work was supported by Consejo Nacional de Investigaciones Científicas y Técnicas de la República Argentina (CONICET) by PIP Nº 0374. M.G.S. thanks support from Consejo de Investigaciones de la Universidad Nacional de Rosario (CIUNR). We acknowledge the financial support of European Commission by the H2020-MSCA RISE projects MELON (Grant N° 872631).

**Data and code availability**

Data will be made available on request.

**Declaration of generative AI in scientific writing**

During the preparation of this work, the authors used ChatGPT (Chat Generative Pre-trained Transformer) in order to enhance the manuscript's readability and writing quality. After using this tool, the authors reviewed and edited the content as needed and take full responsibility for the content of the publication.

**Declaration of competing interest**

The authors declare that they have no known competing financial interests or personal relationships that could have appeared to influence the work reported in this paper.

**Author Contributions**

**L. Imhoff:** writing, data curation, electric and magnetic measurements. **M. B. Di Marco:** sample preparation, optical measurements. **C. Lavado:** DFT calculations. **S. Barolin:** XRD acquisition and analysis, AFM images, review & editing. **M. Stachiotti:** writing, conceptualization, supervision.